# Raman Scattering at Pure Graphene Zigzag Edges

*Benjamin Krauss[1], Péter Nemes-Incze[2], Viera Skakalova[1], László P. Biro[2], Klaus von Klitzing[1], and*

*Jurgen H. Smet[1,*]*

[1] Max-Planck-Institute for Solid State Research, Heisenbergstrasse 1, 70569 Stuttgart, Germany

[2] Research Institute for Technical Physics and Materials Science, H-1525 Budapest, P.O. Box 49, Hungary

*To whom correspondence should be addressed. E-mail: J.Smet@fkf.mpg.de

Theory has predicted rich and very distinct physics for graphene devices with boundaries that follow either the armchair or zigzag crystallographic directions. A prerequisite to disclose this physics in experiment is to be able to produce devices with boundaries of pure chirality. Exfoliated flakes frequently exhibit corners with an odd multiple of 30°, which raised expectations that their boundaries follow pure zigzag and armchair directions. The predicted Raman behavior at such crystallographic edges however failed to confirm pure edge chirality. Here, we perform confocal Raman spectroscopy on hexagonal holes obtained after the anisotropic etching of prepatterned pits using carbothermal decomposition of $SiO_2$. The boundaries of the hexagonal holes are aligned along the zigzag crystallographic direction and leave hardly any signature in the Raman map indicating unprecedented purity of the edge chirality. This work offers the first opportunity to experimentally confirm the validity of the Raman theory for graphene edges.

The electronic states associated with a graphene edge have been the focus of intense theoretical research[1,2,3,4,5] even before the experimental isolation of graphene[6,7]. The edge is either formed by carbon atoms arranged in the zigzag or armchair configuration as displayed in Figure 1a. Zigzag edges are composed of carbon atoms that all belong to one and the same sublattice, whereas the armchair edge contains carbon atoms from either sublattice. This distinction has profound consequences for the electronic properties of the edge states. For instance a ribbon terminated on either side with a zigzag edge has an almost flat energy band at the Dirac point giving rise to a large peak in the density of states. The charge density for theses states is strongly localized on the zigzag edge sites[8]. Such localized states are entirely absent for a ribbon with armchair boundaries. A plethora of different effects associated with the distinct electronic structure of these graphene edges has been predicted by theory including an anomalous quantum Hall effect[9], superconductivity[10] and magnetism[11]. Devices with pure edge chirality to exploit the specific properties of each edge configuration have been put forward. Armchair devices have been proposed as particularly suitable candidates for spin quantum bits[12]. They may offer long coherence times because of the lifted valley degeneracy and the convenient coupling between qubits via Heisenberg exchange. For zigzag ribbons electrostatically controllable valley filter and valves have been dreamed up as devices exploiting the unique features of graphene[13].

To unlock this physics at the edge of graphene, one should first be able to produce devices that possess boundaries with a pure edge chirality and to identify that they have high chiral purity. The observation that mechanically exfoliated flakes frequently exhibit corners with angles that are an odd multiple of 30° initially raised hopes in the community that one edge at such a corner is of the pure zigzag type, while the other possesses the armchair configuration. Atomic resolution scanning tunneling microscopy (STM) and transmission electron microscopy (TEM) at first sight seem predestined to demonstrate that this statement is correct. In practice however TEM at the edge is too invasive[14]. The edge is modified in situ and becomes decorated with unintentional dirt. STM was successfully used to produce and visualize edges on highly oriented pyrolitic graphite (HOPG)[15]. However these edges are not of pure chirality and since flakes are commonly produced on an insulating $SiO_2$ layer STM is hampered. Inelastic light scattering has been put forward as a potential technique to unequivocally distinguish clean armchair and zigzag edges.[16]. The so-called defect or D peak serves as the litmus test. This peak originates from a double resonance process[17]. One of the possible processes[18] is elucidated in Figure 1b and c in momentum space. An electron-hole pair is created (illustrated by the green arrow) by an incoming photon with energy $\hbar\omega_{in}$ in one of the valleys located at the K-point (or K') of the Brillouin zone boundary. The electron (or hole) is then inelastically scattered by a large momentum ($\vec{q}$) zone boundary phonon (black arrow) to an inequivalent Dirac valley at the K'-point (or K-point). An elastic backscattering event returns the electron (or hole) to the original valley, where it completes its Raman roundtrip transition by recombining with its companion hole (or electron) in the course of emitting Raman light at frequency $\omega_{out}$. In view of the small photon momentum, Raman emission occurs only if the elastic backscattering process involves a momentum transfer equal to $-\vec{q}$ (both in absolute value and direction) in order to fulfill overall momentum conservation. This can not be accomplished by a zigzag edge. Along the crystallographic edge direction, momentum remains conserved. Backscattering can only proceed in a direction perpendicular to the edge. For a zigzag edge the momentum can only be transferred in a direction $\vec{d_z}$ which does not allow the electron to return to the original valley in reciprocal space (Figure 1a and c red arrow). Conversely, an armchair edge can convey momentum in the proper direction ($\vec{d_a}$). Summa summarum, only an armchair edge would contribute to the Raman D peak. A zigzag edge would remain invisible in the D peak.

Several Raman studies were reported on flakes exhibiting corners that are odd multiples of 30°. In all cases, the Raman defect line (D peak) from both edges showed similar intensities. The disparity was less than a factor of 2 and the D peak certainly did not vanish for one of the sides as expected and predicted by theory.[18,19,20,21] An example is shown in Figure 1d. Here the Raman measurements were performed with a scanning confocal setup using a solid state laser with a wavelength of 488 nm and an intensity of 7 mW focused to a diffraction-limited spot size of approximately 400 nm. To record Raman maps, the position of the laser spot remains fixed and the sample is raster scanned on a grid with a step size of 200 nm. At each position the backscattered light is dispersed in a monochromator and detected with a Peltier cooled charge coupled device (CCD) with an accumulation time of 1 s. This short accumulation time ensures that no laser induced increase of the D peak occurs due to the gradual disassembly of the graphene flake[22]. Such inadvertent laser induced modifications of the graphene flake are excluded by comparing Raman spectra recorded before and after scanning the sample. The incident laser light can be linearly or circularly polarized. For linearly polarized light the matrix element for the edge assisted Raman transition is maximum when the polarization vector is aligned with the edge and falls as $\cos^2 \theta$ when it is at angle $\theta$. To avoid this angular dependence and the need for adjustment of the polarization direction, the experiments were carried out with circularly polarized light. From these experiments on mechanically exfoliated flakes either one of two conclusions must be drawn: the theory on the inelastic light scattering at the graphene edge is flawed or neither of the edges microscopically consists of pure zigzag chirality even though the average direction aligns with the zigzag crystallographic orientation. Most likely, the second scenario holds, and both edges are composed of a mixture of both zigzag and armchair sections[19]. Note that there are no geometrical constraints which would prevent forming an edge solely out of armchair terminated sections with a different orientation so that on average the edge follows the zigzag crystallographic direction.

This setback has stimulated the search for anisotropic etching procedures[23] that rely on the distinct chemical stability and reactivity of both edge types. Recently two such techniques have emerged. The first method relies on the dissociation of carbon exposed at the graphene edge into Ni nanoparticles, which subsequently act as catalysts for the hydrogenation of carbon at a temperature of approximately 1000°C[24,25,26]. This procedure cuts a network of trenches with a width equal to the metallic particle size. This network is random since it is not possible to guide the nanoparticles along specific trajectories. In the case of graphene, up to 98% of these trenches are at angles of an even multiple of 30° suggesting that these align nearly all along equivalent crystallographic directions of either the zigzag or the armchair type. A second method is based on the carbo-thermal reduction of $SiO_2$ to $SiO$ which consumes carbon from the edge in the process[27]. The reaction is done in argon at a temperature around 700°C. It converts unintentional defects or prepatterned round holes into hexagons all of which have their sides aligned along the same crystallographic orientation. The orientation of the flake was verified using atomic resolution STM images away from the edge but in the vicinity of the etched holes and the edge direction was confirmed to be along the zigzag direction. An example of these hexagons is displayed schematically in Figure 2c and an atomic force microscopy image (AFM) is depicted in Figure 3d. From these results one can conclude that carbon atoms forming an armchair edge have a higher reactivity rate under these experimental conditions and hence boundaries with carbon atoms arranged along the zigzag crystal orientation remain. Further details of the sample fabrication are deferred to the Supporting Information.

The Raman investigations were carried out on samples prepared with this last method producing hexagonal holes. For the sake of comparison we also examine edges of round holes in graphene obtained under conditions where etching is isotropic (see also Supporting Information). For round holes such edges consist inevitably of a mixture of armchair (blue) and zigzag (red) sections as depicted in a cartoon-like fashion in Figure 2a. Figure 3 displays Raman maps of the D (c and f) and G peak intensity (b and e) for the round (top panels) and hexagonal holes (bottom panels). The G peak associated with the zone-center in-plane stretching eigenmode[28],[29] reveals $sp^2$ carbon-carbon bonds. White and black corresponds to high and zero intensity, respectively. Obviously the intensity is low at the round and hexagonal holes as can be verified by comparing with the AFM images on the left. The intensity does not vanish because the diffraction limited laser spot is comparable in size with the etched holes. The D peak intensity is large near the round holes (Figure 3c). The important result can be seen in the D peak intensity map of the sample with hexagonal holes (Figure 3f). The intensity is homogeneous across the sample, and no maxima appear near the hexagonal holes as it is the case for round holes. Figure 4 compares the full Raman spectrum recorded at the round hole marked with the red square in Figure 3a with the spectrum obtained from the hexagonal hole demarcated in blue in Figure 3d. As the intensity of all Raman peaks also depends on, e.g., the amount of graphene probed, laser intensity etc., it is common not to look at the absolute intensity but rather at the ratio of two peaks. Here we focus on the ratio of the D to the G peak intensity I(D)/I(G). The G peak intensity is normalized to 1 in this and all other Raman spectra. The D peak intensity then reflects immediately the ratio I(D)/I(G). From Figure 4b and f it is obvious that the D peak intensity for the hexagonal hole is 1 order of magnitude smaller than that for the round hole (26%). Some statistics are collected in Figure 5. It displays the measured ratio for seven round holes (red region) and for the seven hexagonal holes displayed in Figure 3e (blue region). Also included in the plot is the D peak intensity for regions without holes (bulk). It is not zero but approximately equal to 0.02 (see Figure 4 c and g). We attribute this to some imperfections generated during the preparation of the sample[30]. This background is also visible in the Raman map in Figure 3 c and f and is indicated in Figure 5 with the dashed black line. The laser beam exposes part of the bulk region, and the exposed surface represents a large fraction in comparison with the one-dimensional edge. Therefore this background should be subtracted from the measured peak intensities. Taking this into account the ratio I(D)/I(G) for the boundaries of the hexagonal holes is up to a factor of 30 smaller than for the edges of round holes. It points to a strong discrimination between the different crystallographic chiralities.

   The attachment of functional groups to graphene may affect the D peak intensity, as these groups can alter the hybridization of carbon atoms from $sp^2$ to $sp^3$ or disrupt the lattice symmetry. Since processing steps for the fabrication of round and hexagonal holes differ, the question rises whether the observed differences in the D peak intensity at the edges originate from a different number or type of functional groups, such as for instance hydrogen or oxygen, attached at the edge. This can however be ruled out, since the Raman cross section at visible light excitation for carbon $sp^3$ bonds is minor compared to the resonance enhancement for C-C sp² bonds. In the literature for instance, Raman studies have been reported for hydrogenated graphene[31]. The attachment of hydrogen atoms caused an increase of the D peak intensity. In this case, hydrogen is not only attached at the edge but at the majority of all carbon atoms making up graphene. Despite this large number of $sp^3$ bonds, the D peak intensity only becomes comparable to the G peak intensity. The area probed in the confocal Raman experiment has a diameter of approximately 1μm. The $sp^3$ bonds at the edge only represent a minute fraction of the total number of $sp^3$ bonds. If hydrogen attachment is restricted to carbon atoms located at the boundaries, it would not be possible to resolve these C-H bonds. The same arguments can be invoked for $sp^3$ bonds involving oxygen. Here we can also eliminate a possible influence of oxygen by carrying out an experiment in which the graphene edges are chemically reduced if oxygen is available using a solution of 1mmol ascorbic acid (AA) in 1l of water[32]. AA is a nontoxic reducing agent and was demonstrated to successfully reduce graphene oxide[33]. In order to prevent oxygen from reattaching to the graphene sample, the measurements were carried out with the sample still kept in solution. The Raman spectra for the round and hexagonal holes after 24 hours of AA treatment are displayed in Figure 4d and h. The spectra were obtained at the same position as for the data recorded in Figure 4b and f. The I(D)/I(G) ratio after this chemical treatment has been included in Figure 5 (measurement 9). No change in the I(D)/I(G) ratio occurred. It proves that the edge geometry (zigzag or armchair) is the dominating parameter and that it is not the functional groups attached at the edge.

In summary, we have demonstrated that hexagonal holes obtained by anisotropic etching are bounded predominantly by zigzag edges which do not contribute to the D peak in Raman spectroscopy. Conversely, the absence of a significant D peak near such edges supports a posteriori the validity of the Raman theory which has been developed for graphene edges but could not be confirmed on the corners of mechanically exfoliated flakes. The fabrication of edges with a clean zigzag configuration represents a powerful additional capability in the graphene toolbox. It may be used as a straightforward technique to identify the crystallographic orientation of graphene flakes. By appropriate prepatterning, hexagons may be arranged so as to form constrictions or one-dimensional channels terminated on either side by pure zigzag edges. Also more advanced low-dimensional structures, such as quantum dots bound exclusively by zigzag edges are conceivable.

ACKNOWLEDGMENT**:** We thank D. H. Chae for useful comments on the manuscript and M. Burghard for the idea to use ascorbic acid to remove possible oxygen contaminants. The work in Hungary was supported by OTKA-NKTH Grant 67793.

**Supporting Information Available**:

Successive steps in the fabrication process of hexagonal holes. This material is available free of charge via the Internet at http://pubs.acs.org.

**Figures and Figure Captions:**

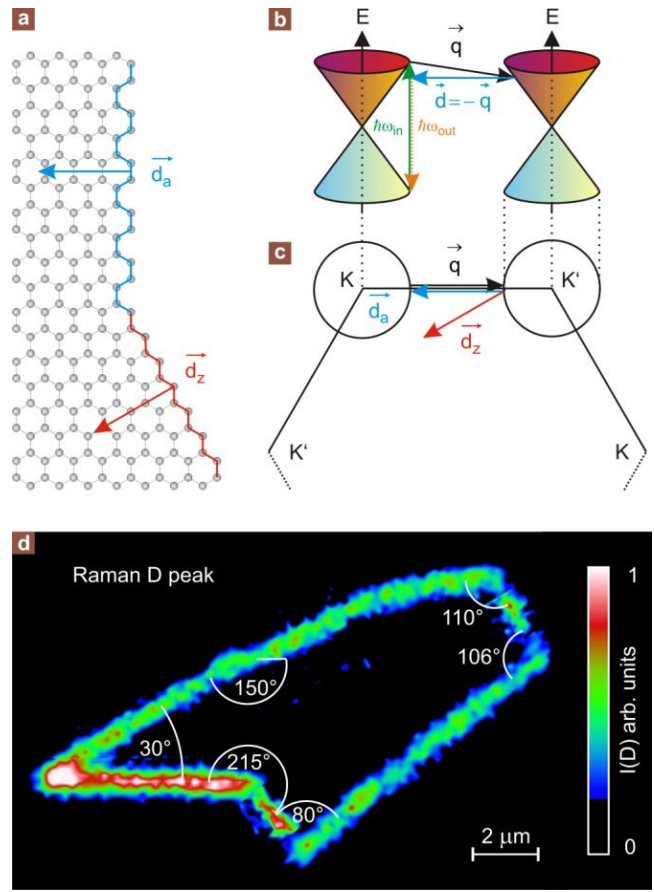

**Figure 1: Raman double resonance mechanism in graphene and at the edge.** (a) Atomic structure of the edge with armchair (blue) and zigzag (red) chirality. The edge can transfer momentum along the defect wave vector $\vec{d_a}$ (blue arrow) and $\vec{d_z}$ (red arrow). (b) Schematic illustration of the double resonance mechanism responsible for the defect induced D peak (see text). (c) First Brillouin zone of graphene and the double resonance mechanism in top view. Only the armchair edge supports elastic intervalley scattering of the electrons or holes. (d) Spatially resolved Raman D peak intensity of a micromechanically cleaved graphene sample. The angles were determined from an AFM image (not shown).

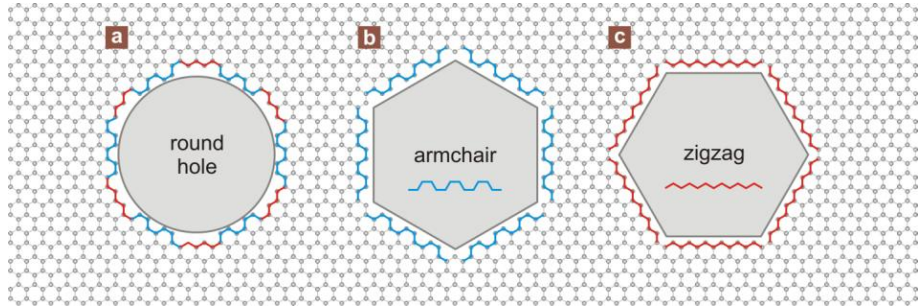

**Figure 2: Lattice model for the edges of round and hexagonal holes. (**a) The edge of a round hole is bound to consist of a mixture of zigzag (red lines) and armchair (blue lines) sections. Holes fabricated by oxygen treatment have rough edges. For the hexagonal holes two cases are possible. Either the circumference is built from armchair segments (b) or from zigzag segments only (c).

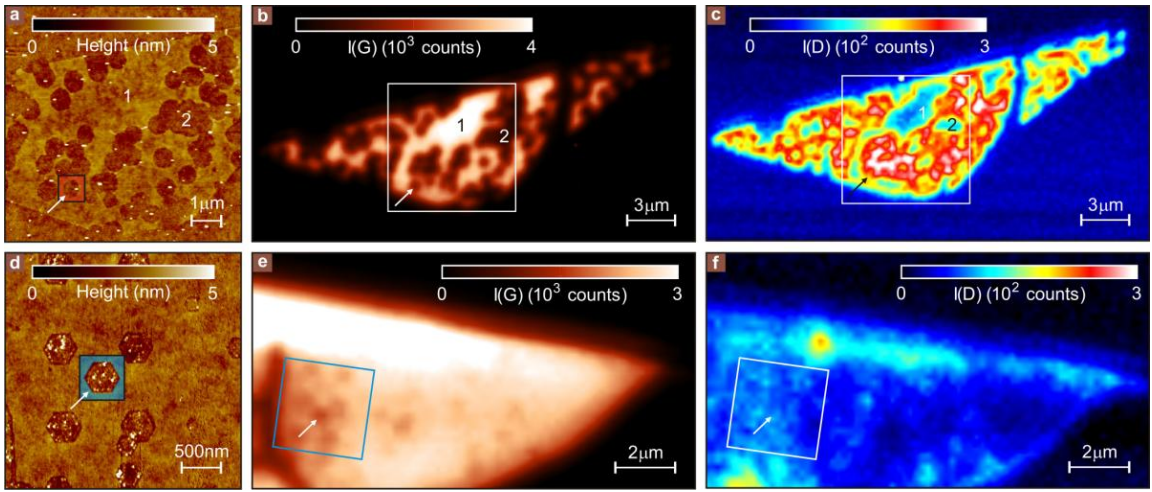

**Figure 3: AFM images and Raman maps of graphene flakes containing round (top panels) or hexagonal (bottom panels) holes. (**a)**,** (d): AFM images of the round and hexagonal holes. (b), (e) Intensity map of the Raman G peak. The G peak intensity is uniform across each flake except at the locations of the holes. These holes appear black (no graphene). The region where the AFM image was taken has been demarcated by a square. (c), (f) Intensity map of the disorder induced D peak. The D peak intensity is high in the vicinity of round holes (c). On the contrary the D peak intensity is not enhanced near the hexagonal holes in **(**f).

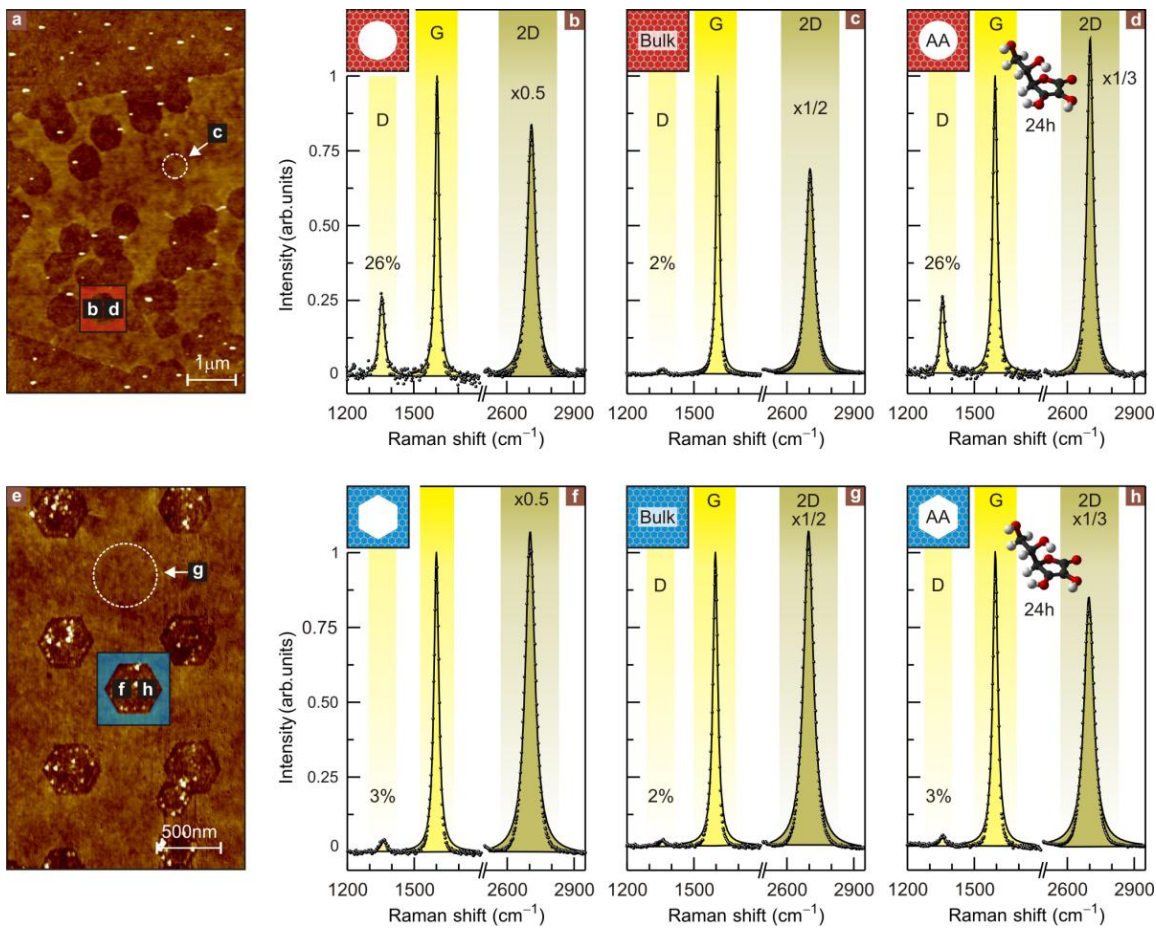

**Figure 4: Raman spectra obtained from a sample with round holes (upper panel) and hexagonal holes (lower panel).** (a), (e) AFM images demarcating the positions where the Raman spectra on the right hand side were obtained. (b) The Raman spectrum of a round hole has a strong D peak whereas the D peak for a hexagonal hole (f) is minimal and only slightly higher than the surrounding bulk value (c), (g). The ratio of the D to G peak intensities is included in percentage for each spectrum. (d), (h) The influence of attached functional groups was ruled out by reducing the samples with ascorbic acid. The Raman spectra resemble the ones before treatment.

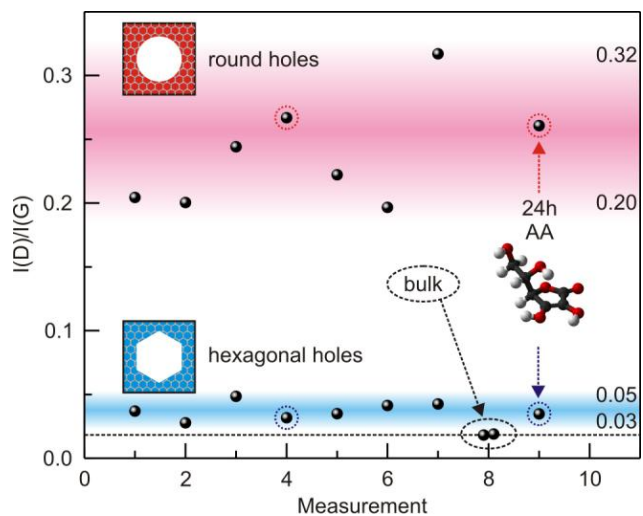

**Figure 5: Statistics for the I(D)/I(G) intensity ratio.** I(D)/I(G) for seven round holes and the seven hexagonal holes in the AFM image in Figure 4e. The encircled values (measurement 4) are obtained from to the Raman spectra displayed in Figure 4 b and f. The dashed horizontal line is the average D peak intensity in the bulk of the sample (0.02, measurement 8 and Figure 4c and g.) and can be considered as a background. Measurement 9 was taken on the same holes as measurement 4 after treating the samples for 24 hours in ascorbic acid (Figure 4 d and h).

REFERENCES

[1] Son, Y. W., Cohen, M. L. & Louie, S. G. *Phys. Rev. Lett.* **2006**, 97, 216803.

[2] Han, M. Y., Özyilmaz, B., Zhang, Y. & Kim, P. *Phys. Rev. Lett.* **2007**, 98, 206805.

[3] Yang, L., Cohen, M. L. & Louie, S. G. *Nano Lett.* **2007**, 7, 3112.

[4] Son, Y. W., Cohen, M. L. & Louie, S. G. *Nature* **2006**, 444, 347.

[5] Moghaddam, A. G. & Zareyan, M. *Appl. Phys. A: Mater. Sci. Process.* **2007**, 89, 579.

[6] Nakada, K., Fujita, M., Dresselhaus, G. & Dresselhaus, M. S. *Phys. Rev. B: Condens. Matter Mater. Phys.* **1996**, 54, 17954.

[7] Wakabayashi, K., Fujita, M., Ajiki, H. & Sigrist, M. *Phys. Rev. B: Condens. Matter Mater. Phys.* **1999**, 59, 8271.

[8] Kobayashi, Y., Kusakabe, K., Fukui, K. & Enoki, T. *Phys. E (Amsterdam, Neth.)* **2006**, 34, 678-681.

[9] Abanin, D. A., Lee, P. A. & Levitov, L. S. *Solid State Commun.* **2007**, 143, 77.

[10] Sasaki, K., Jiang, J., Saito, R., Onari, S. & Tanaka, Y. *J. Phys. Soc. Jpn.* **2007**, 76, 033702.

[11] Yazyev, O. V. *Rep. Prog. Phys.* **2010**, 73, 056501.

[12] Trauzettel, B., Bulaev, D. V., Loss, D. & Burkard, G. *Nat. Phys.* **2007**, 3, 192-196.

[13] Rycerz, A., Tworzydlo, J. & Beenakker, C. W. J. *Nat. Phys.* **2007**, 3, 172-175.

[14] Girit, C. O., et al. *Science* **2009**, 323, 1705.

[15] Tapasztó, L., Dobrik, G., Lambin, Ph., & Biró, L. P. *Nat. Nanotechnol.* **2008**, 3, 397.

[16] Cancado, L. G., Pimenta, M. A., Neves, B. R. A., Dantas, M. S. S. & Jorio, A. *Phys. Rev. Lett.* **2004**, 93, 247401.

[17] Thomsen C. & Reich, S. Double resonant Raman scattering in graphite. *Phys. Rev. Lett.* **2000,** 85, 5214.

[18] Basko, D. M. *Phys. Rev. B: Condens. Matter Mater. Phys.* 2009, 79, 205428.

[19] Gupta, A. K., Russin, T. J., Gutiérrez, H. R. & Eklund, P. C. *ACS Nano* **2009**, 3, 45.

[20] You, Y., Ni, Z., Yu, T. & Shen, Z. *Appl. Phys. Lett.* **2008,** 93, 163112.

[21] Casiraghi, C., et al. *Nano Lett.* **2009**, 9, 1433.

[22] Krauss, B., et al. *Phys. Rev. B: Condens. Matter Mater. Phys.* **2009**, 79, 165428.

[23] Biró. L. P. & Lambin Ph. *Carbon* **2010**, 48, 2677

[24] Datta, S. S., Strachan, D. R., Khamis, S. M. & Johnson, A. T. C. *Nano Lett.* **2008**, 8, 1912.

[25] Ci, L., et al. *Nano Res.* **2008**, 1, 116.

[26] Campos, L. C., Manfrinato, V. R., Sanchez-Yamagishi, J. D., Kong, J. & Jarillo-Herrero, P. *Nano Lett.* **2009,** 9, 2600.

[27] Nemes-Incze, P., Magda, G., Kamarás, K. & Biró, L. P. *Nano Res.* **2010**, 3, 110.

[28] Reich, S. & Thomsen, C. *Phil. Trans. R. Soc. B* **2004**, 362, 2271.

[29] Ferrari, A. C. et al. Phys. Rev. Lett. **2006**, 97, 187401.

[30] Liu, L., et al. *Nano Lett.* **2008**, 8, 1965.

[31] Elias, D. C., et al. *Science* **2009**, 323, 610.

[32] Burghard, Marko, Max-Planck-Institute for Solid State Research, Stuttgart, Germany. Private communication, 2010.

[33] Zhang, J., et al. *Chem. Commun.* **2010**, 46, 1112.